\documentclass[10pt,conference,final]{IEEEtran}
\usepackage{cite}
\usepackage{amsmath,amssymb,amsfonts}
\usepackage{algorithmic}
\usepackage{graphicx}
\usepackage{subcaption}
\usepackage{textcomp}
\usepackage{xcolor}
\usepackage{enumitem}
\usepackage{float}
\usepackage{tikz}
\usepackage{url}
\usepackage{booktabs}
\usepackage{xcolor}
\usepackage{colortbl}
\usetikzlibrary{decorations.pathmorphing}
\usepackage{hyperref}
\usepackage{multirow}

\def\BibTeX{{\rm B\kern-.05em{\sc i\kern-.025em b}\kern-.08em
    T\kern-.1667em\lower.7ex\hbox{E}\kern-.125emX}}
\begin{document}

\title{
Why AI Agents Still Need You: Findings from Developer-Agent Collaborations in the Wild
}

\author{  
\IEEEauthorblockN{  
Aayush Kumar\IEEEauthorrefmark{1},  
Yasharth Bajpai\IEEEauthorrefmark{1},  
Sumit Gulwani\IEEEauthorrefmark{2},  
Gustavo Soares\IEEEauthorrefmark{2},  
Emerson Murphy-Hill\IEEEauthorrefmark{3}  
}  
\IEEEauthorblockA{
\IEEEauthorrefmark{1}Microsoft, Bengaluru, India;  
    \IEEEauthorrefmark{2}Microsoft, Redmond, WA, USA;  
    \IEEEauthorrefmark{3}Microsoft, Sunnyvale, CA, USA\\  
\href{mailto:t-aaykumar@microsoft.com}{\{t-aaykumar}, 
\href{mailto:ybajpai@microsoft.com}{ybajpai},  \href{mailto:sumitg@microsoft.com}{sumitg},  \href{mailto:gustavo.soares@microsoft.com}{gustavo.soares},  \href{mailto:emerson.rex@microsoft.com}{emerson.rex\}@microsoft.com}}  
}  

\maketitle


\begin{abstract}
Software Engineering Agents (SWE agents) 
can autonomously perform development tasks on benchmarks like SWE Bench, 
but still face challenges when tackling complex and ambiguous 
real-world tasks. 
Consequently, SWE agents are often designed to allow interactivity with developers, enabling collaborative problem-solving. 
To understand how developers collaborate with SWE agents and the barriers they face in such interactions,
we observed 19 developers using an in-IDE agent to resolve 33 open issues 
in repositories to which they had previously contributed. 
Participants successfully resolved about half of these issues, 
with those solving issues incrementally having greater success
than those using a one-shot approach. 
Participants who actively collaborated with the agent and iterated on its 
outputs were also more successful, 
though they faced challenges in trusting the agent's responses and collaborating on debugging and testing. 
Our findings suggest that to facilitate successful collaborations, both SWE agents and developers should actively contribute to tasks throughout all stages of the software development process. SWE agents can enable this by challenging and engaging in discussions with developers, rather than being conclusive or sycophantic.

\end{abstract}

\begin{IEEEkeywords}
Agent-based systems, software development tools, collaboration, communication, developer experience
\end{IEEEkeywords}

\section{Introduction}

Software developers are increasingly incorporating generative artificial intelligence (AI) tools into their work to boost productivity\cite{stack_overflow_developer_survey_2024}. Initially, tools like inline code completions and chat assistants have aided developers by generating code snippets and providing on-demand answers. More recently, there has been a surge in the adoption of \textit{software engineering agents} (SWE agents)—tools specifically designed to autonomously perform complex development tasks with the capability to iteratively refine their outputs based on environmental feedback.  They not only generate code, but can also execute tool calls and act on these tools' results~\cite{liu2024largelanguagemodelbasedagents}.

SWE agents have shown promising results on established benchmarks of software development tasks, such as fully autonomously solving many of the GitHub issues in SWE-Bench~\cite{jimenez2024swebenchlanguagemodelsresolve}. However, they still struggle to solve certain issues due to factors such as the complexity and ambiguity of the issues, lack of tacit knowledge, and limited access to the full development environment~\cite{miserendino2025swelancerfrontierllmsearn,Rus_tacitk}. 
To overcome these limitations, many SWE agents are designed to interact with developers, enabling a collaborative approach to problem-solving. This interaction allows developers and agents to share responsibilities such as understanding the codebase, localising, coding, debugging, testing, and reviewing each other's work.

\begin{figure}[H]
    \centering
    \begin{subfigure}[t]{1\columnwidth}
        \centering
        \includegraphics[width=1\columnwidth]{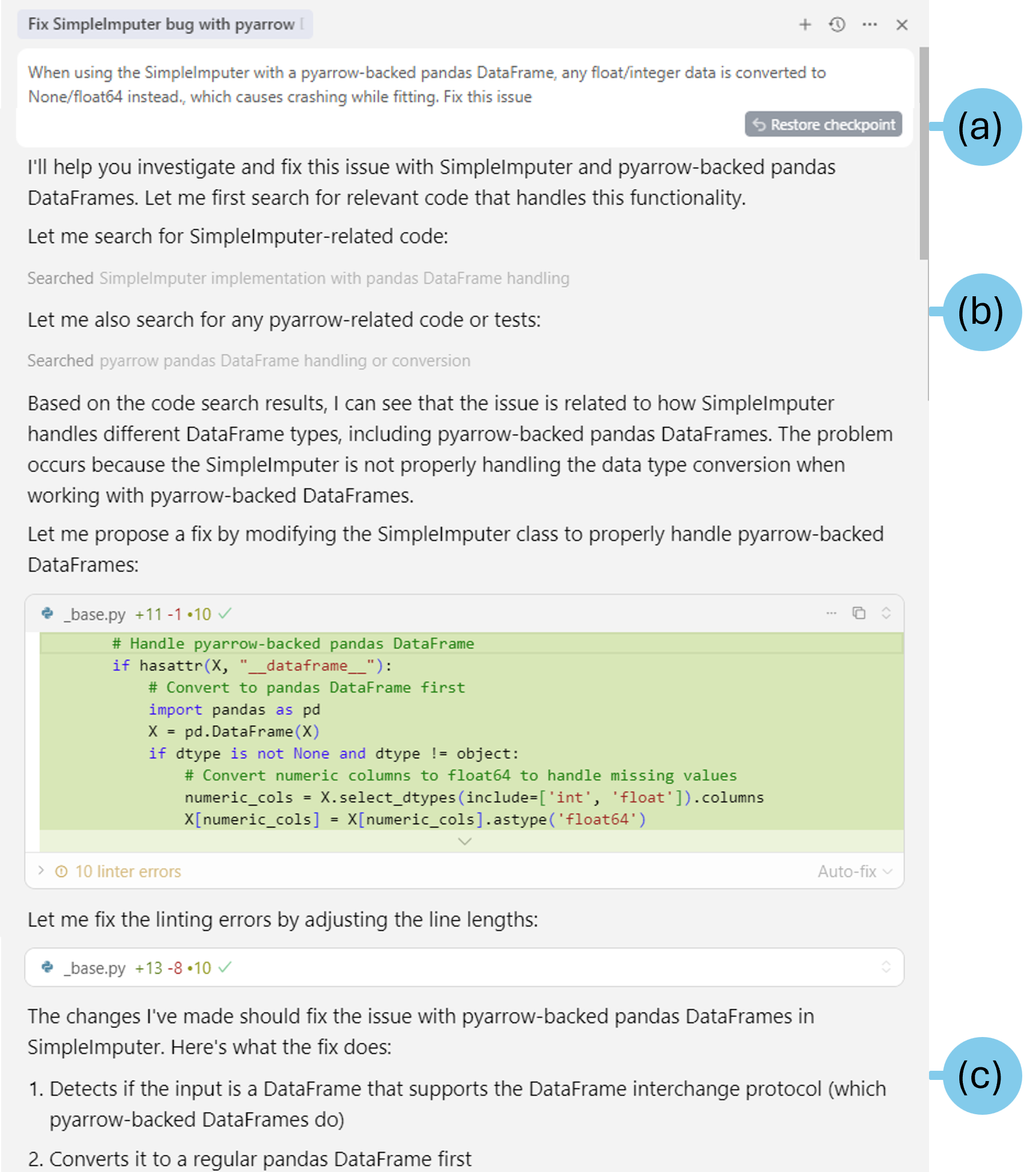}
        \label{fig:screenshot1}
    \end{subfigure}

    \vspace{-4mm}
 
    \begin{subfigure}[t]{0.995\columnwidth}
        \centering
        \includegraphics[width=0.995\columnwidth]{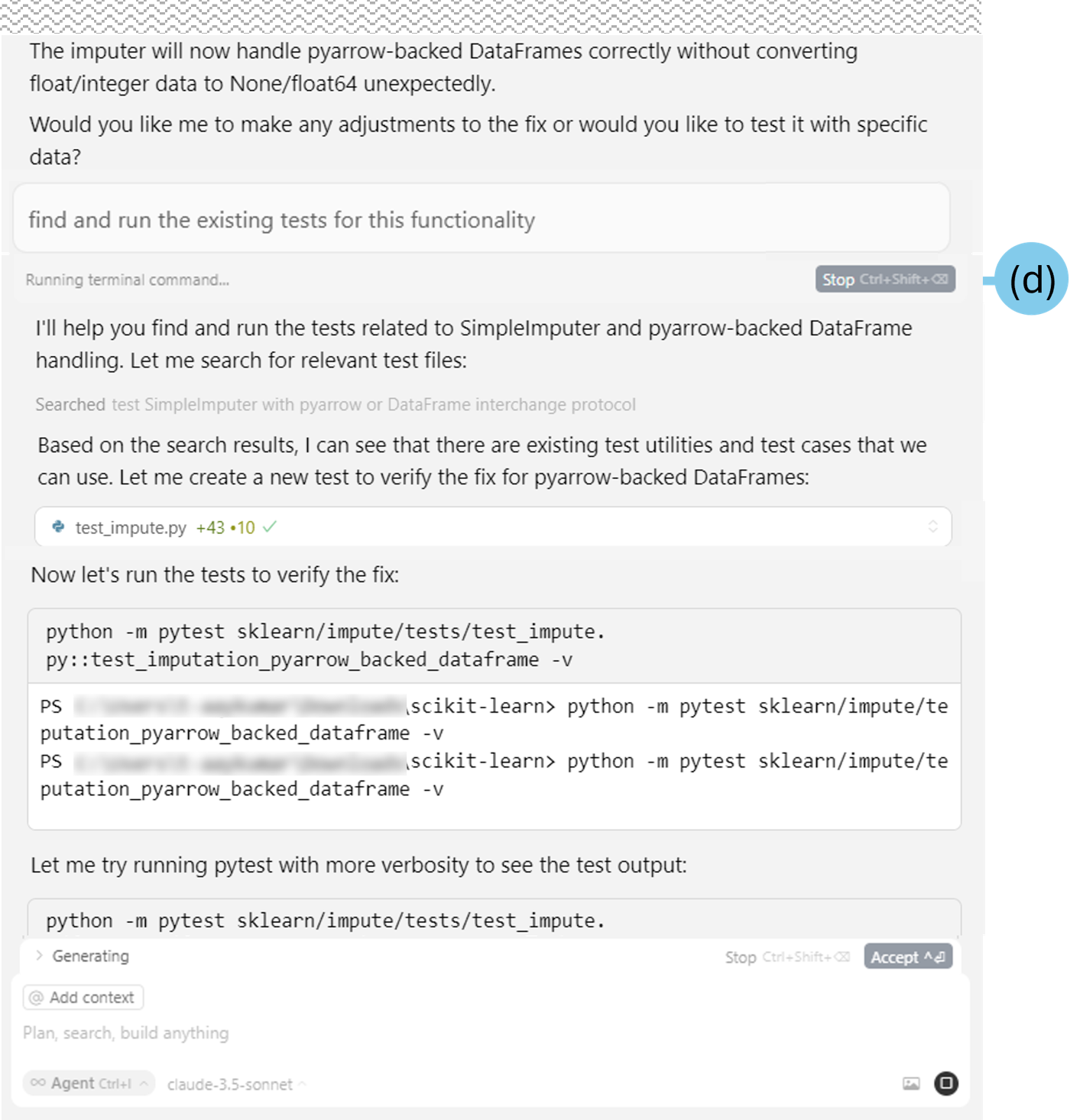}
        \label{fig:screenshot2}
    \end{subfigure}

\vspace{-4mm}

    \caption{Example usage of Cursor Agent (not from our study)}
    \label{fig:workflow}
\end{figure}

\begin{figure}[t]
    \centering
    \includegraphics[width=0.9\columnwidth]{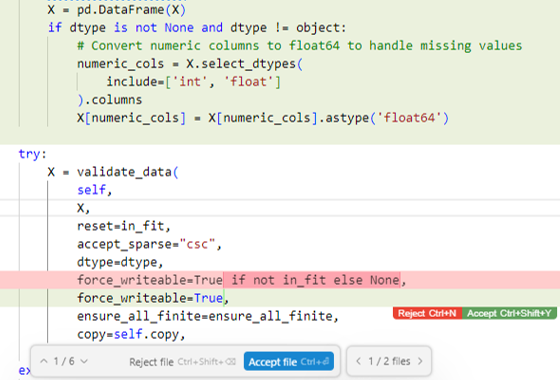}
    \caption{A suggested code edit diff by the agent.}
    \label{fig:additional}
    \vspace{-0.4cm}
\end{figure}

Figure~\ref{fig:workflow} shows an example of one such SWE agent -- Cursor Agent
in the Cursor integrated development environment (IDE) -- responding to a developer's prompt that has been copied from a GitHub bug report \cite{scikit-learn-issue-31373} (this specific bug was chosen purely to illustrate the usage of in-IDE SWE agents and was not part of our study). In the top part of the figure, the agent searches the codebase, describes the issue based on this search, and tries to implement a fix. After prompting, the agent then adds a unit test and runs it. While reviewing the agent's changes, the developer notices a functional call
erroneously changed by the agent (Figure~\ref{fig:additional}, bottom), and
rejects that part of the change, before reviewing the remaining 5 changes.

There is reason to believe that collaborations between agents and developers 
will not be smooth and painless.
For non-agentic tools, prior research has found that 
developers sometimes struggle with understanding and trusting the AI's output, 
hindering effective collaboration~\cite{aiattitudes, groundedcopilot, chilbw2022}.
Is this also the case for SWE agents?
On one hand, unlike non-agentic tools that generate a single response and thus act as a black box, agents can provide developers with information about their process by showing their work as they go. Further, the capability of agents to make decisions autonomously allows for developers to express higher-level goals rather than low-level instructions.
On the other hand, the broader scope of agents’ decisions and subsequent actions can exacerbate trust issues and place a higher cognitive load on the developer of understanding both the agent's reasoning trajectory as well as the mutations it performs to the codebase.
The answer is yet unclear, due to a lack of research on real-world usage of 
SWE agents~\cite{liu2024largelanguagemodelbasedagents}.

In this paper, we begin to address this gap by observing  
19 open-source software developers use SWE agents to solve 
real-world open issues of their choosing. 
Through this study, we aim to address the following research questions:

\begin{enumerate}[label=\textbf{RQ\arabic*:}, leftmargin=*, align=left]
    \item How do software developers collaborate with SWE agents to resolve open issues in active codebases?
    \item What barriers do developers face when collaborating with SWE agents to resolve open issues in active codebases?
    \item What factors influence participants' success when collaborating 
        with SWE agents to resolve open issues in active codebases?
\end{enumerate}

In answering these questions, this study contributes the first empirical investigation of
developer-agent collaboration in resolving real-world issues.

\noindent

\section{Related Work}



\subsection{Agentic Tools for Software Development}
A single definition of an agentic tool has not yet been clearly established and 
evolves as new tools are introduced~\cite{casper2025aiagentindex}. 
Yang and colleagues state that 
a language model `acts as an agent when it interacts with an environment by iteratively 
taking actions and receiving feedback'~\cite{yang2024sweagentagentcomputerinterfacesenable}, 
while Jin and colleagues mention that the ability to `select 
the optimal solution from multiple homogeneous results' is fundamental to an 
agentic tool~\cite{jin2025llmsllmbasedagentssoftware}. 
Aside from generating textual responses, agents can perform actions such as 
searching for files, modifying code,  
and running terminal commands by calling tools. 
Further, agents can iterate on their own output, enabling them to complete 
tasks without the need of human intervention~\cite{liu2024largelanguagemodelbasedagents}.



In software development, there has been a rapid rise in the number of agentic tools 
in research and in practice~\cite{casper2025aiagentindex}. 
Many of these tools aim to solve end-to-end software engineering tasks, as reflected in the 
evolution of benchmarks such as  
SWE-bench~\cite{jimenez2024swebenchlanguagemodelsresolve}, 
a set of real previously-solved GitHub issues,
and
LiveCodeBench~\cite{jain2024livecodebenchholisticcontaminationfree}, which adds new
issues over time.
Similarly, SWE-Lancer~\cite{miserendino2025swelancerfrontierllmsearn} uses real-world freelance 
software engineering tasks to evaluate AI tools.
While these benchmarks aim to evaluate real-world performance, they all test 
the \textit{autonomous} performance of agentic tools, rather than their performance with 
a human-in-the-loop. 
As Liu and colleagues report, most research-based
agentic systems target maximum automation, where human involvement is limited 
to specification of the problem statement~\cite{liu2024largelanguagemodelbasedagents}. 

At the same time, several agentic tools have been integrated into IDEs, 
such as 
VSCode Agent Mode \cite{vscode_agent_mode}, 
Windsurf Cascade \cite{windsurf_cascade} and 
Cursor Agent \cite{cursor_agent}.
Unlike fully autonomous agentic systems, these tools are designed to operate
with a human-in-the-loop.
Although such tools are rapidly emerging and evolving, at the time of this writing,
we know of no empirical research about how developers use such in-IDE agentic tools.
This paper aims to fill this gap.

\subsection{Developer Experiences with AI Tools}

Prior research has explored the strategies that developers 
apply when using AI~\cite{hai_ide_litrev},
especially those related to prompt engineering~\cite{braberman2024taskspeopleprompttaxonomy, promptstaxo, santu2023telergeneraltaxonomyllm}.
Prior work has also found that several development tasks
are time-consuming or challenging when working with AI developer tools, 
including 
communicating the required context to AI tools~\cite{groundedcopilot, hai_ide_litrev},
verifying AI-generated code recommendations~\cite{cups}, and
debugging incorrect AI-generated code~\cite{chilbw2022}.
Developer studies are mixed about whether AI 
increases developer efficiency~\cite{chilbw2022} or decreases it overall~\cite{copilotstudy}.
Studies have also found that developers often use AI tools as learning aids 
and information sources~\cite{chilbw2022, haque2024informationseekingusingai},
and have concerns about the security~\cite{chilbw2022, copilotstudy}, quality and reliability~\cite{cups, aiattitudes} of generated code. 
While these studies have richly explored developer interactions with 
AI code completion~\cite{cups, chilbw2022, groundedcopilot, paradis2024doesaiimpactdevelopment} and 
AI-based chat~\cite{robin, dayeicse2024,programmersassistant, chattingwithaichatgpt}, there is 
a dearth of developer experience studies on the use of SWE agents.
Yet, SWE agents offer a unique interaction paradigm, 
differing from non-agentic tools in their capability to perform complex tasks 
autonomously and take feedback from the 
environment~\cite{liu2024largelanguagemodelbasedagents}. 


Such capabilities are also found in software bots, 
which can make asynchronous edits to codebases. 
Prior developer experience research has found that these tools 
sometimes perform unsolicited actions and produce too much information~\cite{botsinse, ossbots}.
Largely pre-dating the current wave of LLM-based AI,
such bots typically operate on strict instructions;
in contrast, SWE agents can make decisions autonomously, 
potentially complicating effective collaboration~\cite{chopra2024exploringinteractionpatternsdebugging}.

Prior work has argued for the importance of building design 
guidelines for agents to facilitate effective interaction mechanisms~\cite{bansal2024challengeshumanagentcommunication, liu2024largelanguagemodelbasedagents}. 
Recent work has begun to address this gap for SWE agents. 
While Epperson and colleagues discussed the challenges developers face when debugging autonomous multi-agent systems~\cite{steeringmultiagent}, Pu and colleagues studied the effect of proactivity 
in developers' interactions with agentic tools, 
reporting that contextual awareness and salience of the agent's actions are critical to avoid workflow disruptions~\cite{proactive}.
We build on this line of work by studying interactions with SWE agents in
real-world tasks.





\subsection{Communication in Software Development Teams}
Prior work has found that effective communication of knowledge in software 
engineering is a complex problem that requires active involvement from all collaborators. 
Rus and colleagues report that the majority of software engineering 
knowledge is \textit{tacit}, that is, gained through personal experience, 
rather than explicitly documented. 
This can make it difficult for new team members to work effectively, as they 
lack the knowledge that current team members have \cite{Rus_tacitk}. 
Gonçalves and colleagues correspondingly reported that developers 
spent the most amount of time in collaborative work if they were a new team member or if they had business/customer 
knowledge \cite{collabGoncalves}. 
Prior literature further suggests that a lack in streamlined, synchronous communication can result in mistrust and 
frustration among team members, particularly in distributed software development teams \cite{trustvirtualseteam, LENBERG201515BSE}. 
It is thus important for both managers \cite{managerinse} as well as junior 
software developers \cite{juniorsecommunicate} to be effective communicators 
and collaborators.

Kuttal and colleagues \cite{haivshhwizardofozkuttal,pairbuddy} found through Wizard-of-Oz studies that developer-agent interactions tend to suffer due to a lack of extended discussions and the agent not understanding non-verbal cues. Extending this line of work, we aim to leverage findings on communication barriers in software development teams to investigate such barriers in collaborations between developers and in-IDE SWE Agents.
\section{Study Design}

To answer our research questions,
we aimed to observe developers collaborating with a SWE agent
in a highly ecologically valid setting (in `the wild').
To achieve this,
we asked 19 professional developers to use Cursor Agent
to resolve issues in a codebase to which they have previously contributed.
\subsection{Tool Selection}\label{sec:tools}

In selecting the appropriate SWE agent for our study, we considered the spectrum of human-AI collaboration capabilities, ranging from highly autonomous agents to those that enable flexible interactions with developers. We reviewed several agents, starting with fully autonomous options like AutoCodeRover \cite{zhang2024autocoderoverautonomousprogramimprovement}, and moving to agents that allow a moderate level of human-AI collaboration. These moderate collaboration tools often operate outside the IDE and primarily rely on user interaction through prompting, such as Devin \cite{devin}, Aider \cite{aider},OpenHands~\cite{wang2024openhandsopenplatformai} and GitHub Copilot Workspaces \cite{github_copilot_workspace}. Finally, we considered agents that operate within an IDE, facilitating high collaboration not only through prompting but also by providing developers access to the full development environment. Examples include Cursor Agent \cite{cursor_agent}, VSCode Agent Mode \cite{vscode_agent_mode}, Windsurf Cascade \cite{windsurf_cascade}, Cline \cite{cline}, and Amazon Q Developer \cite{amazon_q_dev}.

We decided to focus on IDE-based agents as they offer more flexibility for developers to choose whether to rely fully on the agent or closely collaborate on tasks. Moreover, IDE-based AI tools like GitHub Copilot, Cursor, and Windsurf have gained significant popularity in recent years.

To select among these candidate tools, we assessed the types of features and interaction mechanisms present in each of these tools (as of April 30, 2025). These features included those related to flexibility in interactions (modelessness, ability to preview code changes, ability to backtrack within conversations, ability to synchronously edit the same file as the agent), context used by the agent (files opened in IDE, recent user actions) and the interpretability of the agent's outputs (code change explanations, visualization of agent's reasoning). We found that Amazon Q was lacking some of these features,
such as providing explanations of its outputs.
VSCode Agent Mode was newly released at the time of our study and not yet as mature as other options.
Although Cline is a capable SWE agent, 
the grey literature suggests that it has a 
`steeper learning curve, less polished UI than Cursor or Windsurf' \cite{pahwar2025ai},
making it less appropriate for a study with limited time for participants to learn the tool.

This left Cursor Agent and Windsurf Cascade as our final candidates. Both of these tools contained almost all of the features that we found in our candidates.
We conducted pilot testing using both tools with a small convenience sample of developers
and found the tools comparable in capabilities and usability.
However, Cursor Agent offered one feature that made it 
particularly suitable for our study: automatic incorporation of users'
currently open files and cursor position as context for the agent.
Therefore, we selected Cursor Agent as our study tool. Participants used Cursor v0.47 with Claude 3.5 Sonnet.

\subsection{Tasks}

We sought software engineering tasks for this study that:

\begin{itemize}
    \item reflected developers' typical day-to-day tasks;
    \item participants were motivated to solve thoroughly;
    \item could be completed in less than an hour,
         enabling us to analyze end-to-end agent-participant collaboration; and
    \item would not use proprietary code, which participants may not be allowed
        to show us as researchers or to send to a third-party AI tool.
\end{itemize}

\noindent

Thus, we asked participants to work on recent open issues (such as bugs or feature requests) 
in open-source repositories that they had worked on before. We applied the following guidelines to select suitable repositories that were:
\begin{itemize}
    \item of significant size, that is, with more than 50 
          source code (i.e., non-documentation and non-configuration) files; 
    \item actively maintained (i.e., commits in the prior month);
    \item software centric, thus excluding repositories such as a collection of study materials or                                 tutorials~\cite{10.1145/3671016.3671401}; and
    \item starred over 500 times on GitHub, an 
          indicator of the maturity of the repository~\cite{sens2025largescalestudymodelintegration}.
\end{itemize}
\noindent
Further, we limited our selection to open-source repositories that were part of two major
GitHub organizations that are managed by our company. 
This enabled us to select participants based
on internal-only information about repository contributors, like 
seniority and geographic location, and to communicate
more easily with those contributors.

While most participants selected their first task based on the list 
of open issues in their project,
some (five) participants selected tasks to work on before the start of the session, some of which were not GitHub issues. We allowed them to continue as they had planned.
We further asked participants to choose issues that could be 
solved based on code edits, rather than configuration or documentation fixes, 
to ensure the agent could provide meaningful assistance. 

If participants had time to work on multiple tasks during the session, 
after they completed their first task,
we collaborated with them to choose subsequent issues 
to ensure diversity in task type (e.g., if the first issue was a bug, 
    we encouraged participants to next select a feature request) and difficulty (e.g., if the first issue was completed 
    easily, we encouraged participants to next select an issue that seemed more challenging).

\begin{table}[t]
\caption{Participant Demographics}
\centering
\begin{tabular}{@{}p{1.3cm}p{7.1cm}@{}}
\toprule
\textbf{Dimension} & \textbf{Details} \\
\midrule
\rowcolor{gray!5}Professional Role & Software Engineer I/II: 6; Senior Software Engineer: 7; Principal Software Engineer/Manager: 5; Associate Consultant: 1 \\ 
\arrayrulecolor{gray!50}\cmidrule{1-2}\arrayrulecolor{black}  
\rowcolor{gray!5}Professional Experience & 0-2 years: 1; 3-5 years: 4; 6-10 years: 6; 11-15 years: 3; \hspace{2em}$\ge$16 years: 5 \\
\arrayrulecolor{gray!50}\cmidrule{1-2}\arrayrulecolor{black}  
\rowcolor{gray!5}Gender & Men: 14; Women: 5 \\  
\arrayrulecolor{gray!50}\cmidrule{1-2}\arrayrulecolor{black}  
\rowcolor{gray!5}Age & 18-25: 3; 26-35: 10; 36-45: 4; 46-55: 2 \\  
\arrayrulecolor{gray!50}\cmidrule{1-2}\arrayrulecolor{black}  
\rowcolor{gray!5}Race & White: 6; South Asian: 5; Asian: 2; Black or African American: 2; American Indian or Alaska Native: 1; Hispanic or Latino: 1; Jewish: 1; Multiracial: 1 \\  
\arrayrulecolor{gray!50}\cmidrule{1-2}\arrayrulecolor{black}  
\rowcolor{gray!5}Region & US: 10; India: 2; Kenya: 2; Israel: 2; Germany: 2; China: 1 \\  
\bottomrule 
\end{tabular}  
\label{tab:demographics}  
\vspace{-0.3cm}
\end{table}

\subsection{Participants}\label{sec:participants}

After shortlisting repositories, we reached out to 
a total of 43 contributors who were employees of our company
and who had contributed at least one commit to the 
repository in the previous 4 months.
We aimed to recruit a diverse sample of participants by
reaching out to participants belonging to different regions, 
genders, levels of seniority and with different levels of activity 
in the repository. 
26 candidates declined or ignored our invitation to participate.
We recruited two participants based on recommendations 
from other invitees who were unavailable. 
In total, 19 participants accepted our invitation and completed the study. Participants demographics are presented in Table~\ref{tab:demographics}.
18 of our participants reported having experience with Github Copilot Chat in their regular programming work, while 3 participants reported using Cursor.

\subsection{Protocol} \label{sec:protocol}

Each study session was a meeting between the participant and 
the first author, conducted using a video conferencing tool. 
Sessions were scheduled for 60 minutes, with some variability based on 
configuration issues and participants' availability beyond the scheduled time. To ensure that such variations would not affect participants' behavior, we informed them that the objective of the study was to observe their interactions with the agent rather than to solve as many issues as possible. Further, we did not analyze time-dependent variables such as action durations and counts, focusing instead on the presence or absence of these actions (as described in Section \ref{sec:analysis}).

At the start of each session, the study administrator asked the participant 
to fill out a pre-study questionnaire and consent form.
Following this, the study administrator gave a tutorial of the Cursor IDE 
and its interface by working through a demonstration task to ensure that 
participants understood the tool, its usage, and its features.
Participants were instructed to think aloud 
and to focus on solving the issue, rather than assessing the AI, 
to create a more realistic environment. 
They were also instructed to try to use the AI for all their tasks, 
and make small manual changes if necessary -- this instruction was added after 
some pilot study participants stopped using the agent altogether after switching
to manual editing.

Participants worked on their tasks by remotely controlling 
the study administrator's screen.
We performed all the necessary set up for the participant's 
repository on the study administrator's system. 
We chose not to use the participants’ systems to eliminate the
burden on participants on having to setup the Cursor IDE with 
their codebase.
However, in one case, the participant worked 
on their own computer because 
there were configuration issues with the setup on the study administrator's system
and because the participant had previously installed and worked on the Cursor IDE.

Once participants completed an issue, the study administrator first asked the participant 
if they were satisfied enough with the fix to submit it as a pull request. 
If not, participants were asked to keep working to reach that level of 
satisfaction (if possible within the scope of the session) before moving on to the next issue. 
We did not ask participants to create and submit a pull request since that would 
require them to enter their GitHub credentials onto the study administrator's system. 
Nonetheless, so that participants could create a pull request with their credentials, 
we sent them the diff of their work as a patch after the session. 

At the end of each session, participants filled out a post-study questionnaire 
containing questions about their perceptions of the tool.
\subsection{Data Collection}
We collected several sources of 
data during the study, as described below.

\subsubsection{Session recordings}

To understand participants' usage of the agent in the session, 
we used video, audio, and screen recordings
to qualitatively code both participant actions as well as the details of their 
interactions with the agent. 

\paragraph{Participant Action Codebook} 

We coded participant actions by adapting the 
CUPS taxonomy, previously used for 
understanding common programmer actions when using 
AI code completion~\cite{cups}. 
We modified this taxonomy to add states unique to collaborating 
with SWE agents, 
such as following the trajectory of the agent's execution and 
backtracking to a previous point 
in the conversation (Figure~\ref{fig:workflow}a).  

\paragraph{Chat Trajectory Codebook} 

We also examined the trajectories of the developer-agent 
interactions in our study by 
coding information on the context the participant provides to the
agent (e.g., files, code snippets, and external links) 
as well as the information they gave to and sought from the agent in their prompts. 
To create our codebook, we performed open coding on two participants' traces, then 
iterated on this codebook by annotating trajectories for two further participants 
to create the final version of our codebook.

To validate each of our codebooks, one author coded a study session by applying codes to time segments across the session recording. 
These codes were then masked, and another author recoded the 
same session by applying codes to the corresponding time segments. 
We then computed Cohen's $\kappa$, yielding inter-rater reliability scores of 0.92 and 0.89 for the participant action and chat trajectory codebook respectively, indicating near perfect agreement~\cite{landis1977measurement}.

\subsubsection{Pre-Study Questionnaire}

This questionnaire consisted of optional questions about participant demographics
and prior experience with AI tools.
We asked questions on prior experience to explore how this may be
related to participants' use of the tool~\cite{dayeicse2024}. These questions were based on previous studies of developer-AI interaction~\cite{cups,dayeicse2024}. 
We also included a question about participants' perceptions of AI tools 
for programming in terms of usability~\cite{programmersassistant} 
by adapting the Technology Acceptance Model questionnaire~\cite{tam}. 

\subsubsection{Post-Study Questionnaire}

This questionnaire contained questions about 
the usability of the tool;
the perceived role of the agent in the development process~\cite{programmersassistant}; 
the importance of the design features of the tool~\cite{dayeicse2024}, where we drew features from the tool assessment in Section~\ref{sec:tools}; and 
an open-ended question for participants to describe 
any notable observations while using the tool.

\subsection{Analysis Approach} \label{sec:analysis}

Based on the relevance of actions to the immediate state of the chat with the agent,
we interpret the presence and ordering of participant actions at prompt-level granularity (e.g., verifying code changes, stopping the agent) or issue-level granularity (e.g., writing code, running tests) and analyze codes on chat interactions at the prompt-level granularity. Since participants were instructed to think aloud and study sessions varied in duration, we do not analyze the durations of participants' activities, nor do we analyze the absolute counts of these activities to avoid biases toward participants who performed actions repetitively in shorter intervals. We also note specific utterances and scenarios that indicate barriers to effective collaborations, reporting only those that recur across different participants.

While our results are largely qualitative,
we include some numerical data throughout to give the reader
a fuller sense of the trends that we observed.
However, given the exploratory nature of the study and the 
relatively small sample size, we caution the reader against
broad extrapolation or generalization.
Congruently, we present findings descriptively, 
without applying inferential statistics.

Towards helping to ensure the validity of our findings, 
we sent an early version of our 
results to all participants as a member check~\cite{motulsky2021member}.
Our study protocol has undergone privacy assessment and review by
our organization.
To aid in replication,
qualitative codebooks and 
study materials can be accessed in the Appendix.

\begin{figure*}[ht]
    \centering

    \includegraphics[width=\linewidth]{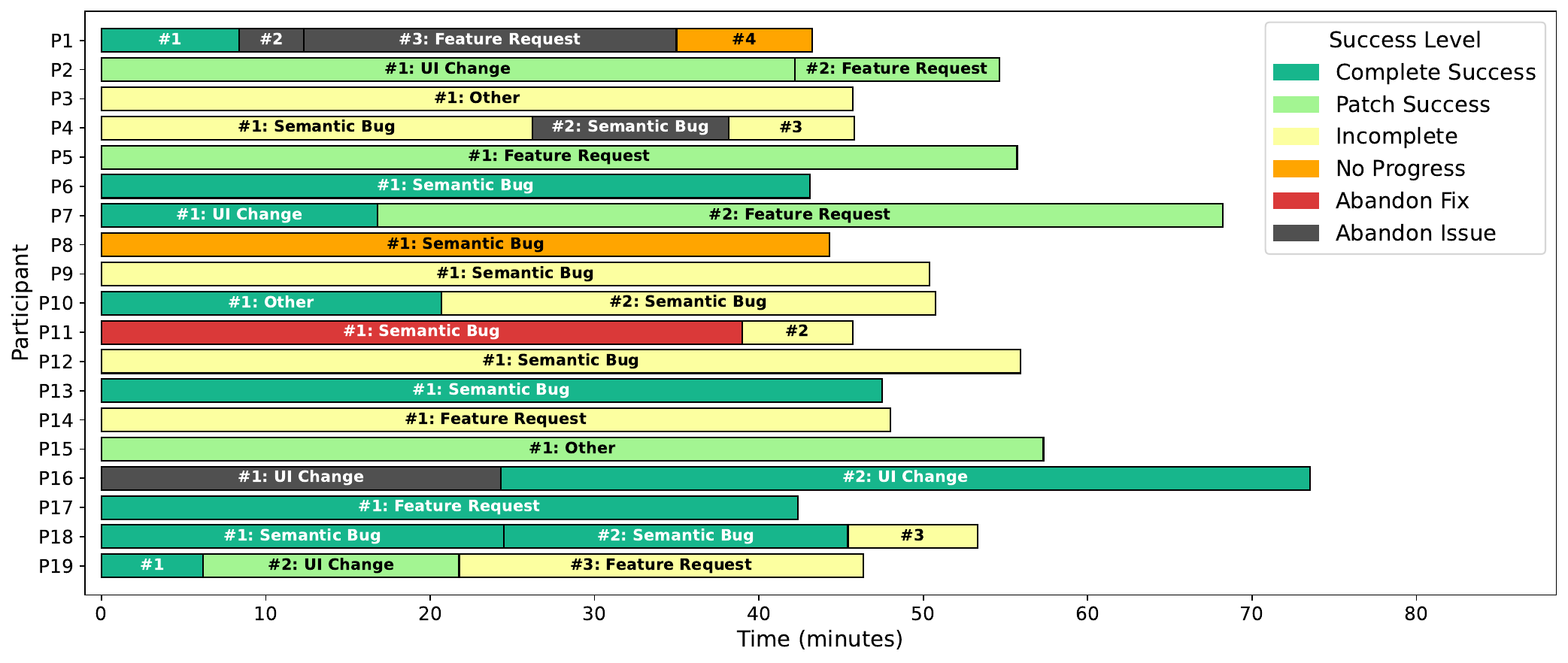}
    \caption{Issue Timeline by Participant with Success Rates. Success Levels are detailed in Section~\ref{greatsuccess}. \label{fig:success}}
    \vspace{-0.1cm}
    \vspace{-0.4cm}
\end{figure*}

\subsection{Limitations}
One limitation of our study is that we recruited participants 
solely from one company. 
Participants' interactions and perceptions of the tool are likely 
influenced by the corporate culture and organizational 
incentives at this company, 
which is actively promoting AI  products and experiences.
Likewise, we acknowledge our positionality as researchers in that company, 
and the inherent unconscious biases and incentives that this environment introduces in our 
interpretation of the study data.

Another limitation is small sample size (19 participants);
we accepted this limitation so that we could perform time-consuming, thorough action
and trajectory coding which does not easily scale up to more participants. 
However, since participants in our study worked on open issues that did not have predefined solutions, we cannot measure or understand the effect of the complexity of these tasks on their performance.
Further, external factors such as participant demographics may have affected the results of our study, though we have no evidence for this: upon performing a logistic regression with success as the dependent variable and the predictors being age, gender, professional role, years of experience and number of commits to the repository in the 4 months leading up to the study; none of the predictors were statistically significant. 

Most participants in our study had not previously 
used Cursor or any other SWE agent, 
and thus our results mostly reflect early usage,  
rather than regular long-term use. 
Our study is also subject to observer effects; 
as participants knew that we were studying their interactions, 
they might may have used the tool differently than in a private setting.
Likewise, our instruction that participants try to use AI for all
their tasks means that, in the wild, developers would likely
be more prone to abandonment. Variations in the duration of study sessions might also have affected participants' behavior. Furthermore, while model call latency was the overall dominant latency bottleneck, our decision to have participants use a remote machine induced an additional UI latency which may have hindered their user experience.

Our choice of Cursor in early 2025 also has an effect on
generalizability; while we found Cursor to have the most interaction features compared to other in-IDE SWE Agents and we anticipate our findings would apply to such tools, our findings may not generalize to non-IDE based SWE Agents.
Further, since AI tools and models are constantly evolving, 
some findings reflect present-day limitations of agents and their underlying models.

\vspace{-2mm}
\section{Results}
\vspace{-1mm}

Figure~\ref{fig:success} visualizes participants' success and time spent as they submitted 269 prompts to the agent across 33 issues. Participants were satisfied with the final patch for 16 of these tasks, while 4 of these tasks were abandoned for external reasons (such as participants 
being unconvinced of the issue's validity) for a total success rate of 55\% $(16/(33-4) )$.
In this section, 
we describe the strategies participants applied,
the challenges they faced while collaborating with the agent,
and finally how these strategies and other factors were associated with participants' success. 






\vspace{-1mm}
\subsection{RQ1: How developers collaborate with SWE agents}
\vspace{-1mm}

In this subsection, we describe the patterns and strategies we observed in participants' collaborations with the agent.

\subsubsection{Delegation Strategies} \label{sec:collab-sat}
We observed that participants broadly followed one of two 
approaches to delegate work --

In one approach (the \textit{one-shot strategy}), 
participants provided the agent with the entire issue description 
and asked it to solve the issue. 
Ten out of 17 participants who worked on GitHub issues followed this strategy. 
This is a high risk, high reward approach: 
\begin{itemize}
    \item If the agent is able to generate a comprehensive fix and verify its changes, the participant can resolve 
          the issue with little additional effort. 
    \item If the agent is unable to generate a comprehensive fix or verify its changes, 
          participants must then manually both understand the generated fix and 
          then iterate on it to get to a valid fix. 
          We observed that this iteration can be time-consuming, 
          as the it can entail not only understanding the code change itself,
          but also understanding the reasoning behind the change, 
          navigating between code changes, 
          and understanding changes to test files. 
\end{itemize}

In the other approach (the \textit{incremental resolution strategy}), 
participants manually divide the task into sequential sub-tasks 
and ask the agent to solve each sub-task sequentially in separate prompts. 
When the participant is satisfied with the solution to one sub-task, 
they then ask the agent to work on the next sub-task. 
Although this approach is safer than the one-shot strategy 
because mistakes made by the agent can be reviewed and caught earlier, 
it requires more proactive involvement and interaction from the 
participant to divide the issue into these sub-tasks.
As evidence,
participants who used this strategy used an average of 11.0 prompts per issue,
compared to 7.0 for the one-shot strategy. 
Further, these participants were more likely to manually 
read existing code to enhance their own understanding 
while working on issues, doing so in 83\% (15 out of 18) of the issues they 
worked on as compared to 60\% (9 out of 15) for other participants.

One participant (P7), who had used Cursor before, 
explained that they preferred different strategies 
in different scenarios: 
\textit{``For an issue that is very contained, 
I trust it to come up with the changes itself, 
but for issues that are more involved with different parts 
of the codebase I feel have to point it in the right direction 
so that anything sensible comes out at the end"}. However, we also observed that these approaches are not mutually exclusive; 
when delegating the entire issue to the agent via the one-shot strategy, 
an incomplete solution may require the participant to prompt the agent with extra steps via incremental resolution. 
As P3 mentioned, \textit{``The AI agent was able to generate good structure... once more context was needed, it committed errors that need more handholding 
and additional time to be spent on guiding it”}.

\begin{figure}[t]
    \centering
    \includegraphics[width=\linewidth]{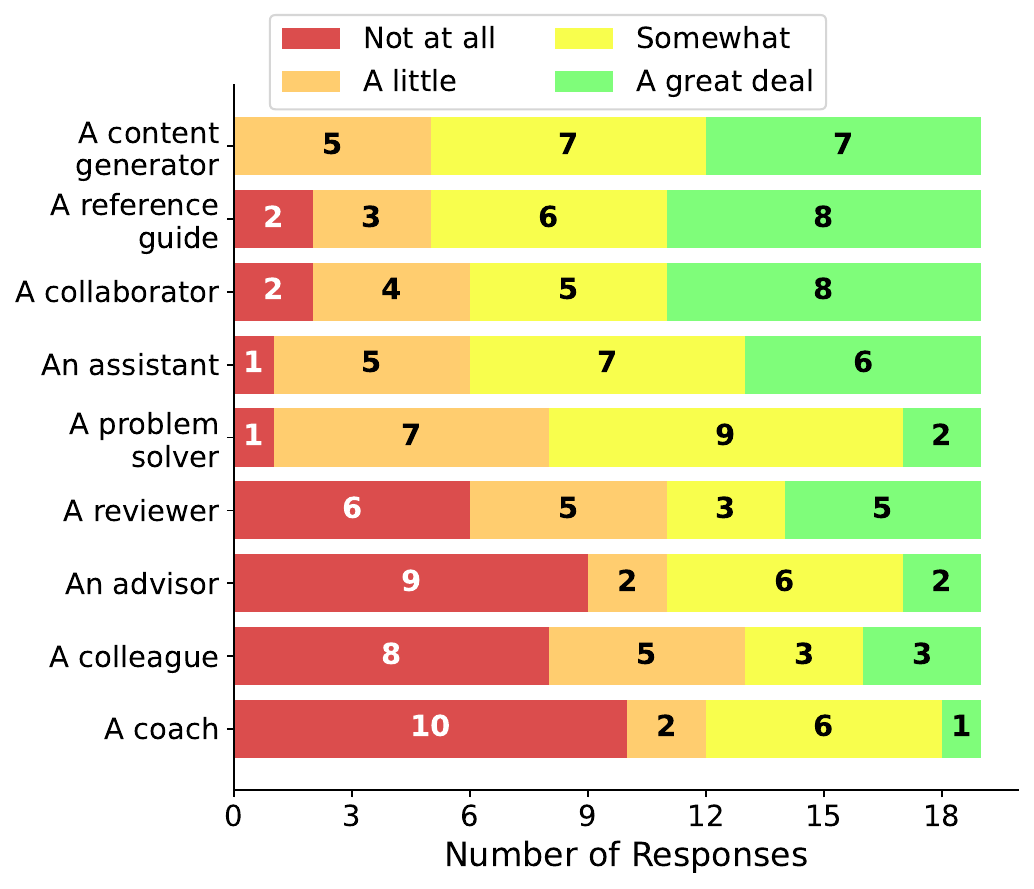}
     \vspace{-0.65cm}
    \caption{Post-study questionnaire responses on Agent Perception}
    \label{fig:perception}
    \vspace{-0.4cm}
\end{figure}

\subsubsection{Providing Expert Information}\label{sec:expert}
We observed that participants provided two different types of information to the agent to help it make the desired code changes:
\begin{itemize}
    \item\textit{Contextual information} from the issue description 
        and the environmental context (such as test logs). Eg: \textit{``There are build errors in your solution"}
    \item\textit{Expert information} relating to code implementation or convention 
        that cannot be observed directly from context or the issue description 
        and is based on the developer's prior knowledge of the repository. Eg: \textit{``Usually we don't set an upperbound. I think 3.10 should be fine now"}
\end{itemize}

Participants' patterns of providing such information to the agent mirror 
those of software engineering managers.
Kalliamvakou and colleagues~\cite{managerinse} report that when assigning work,
great managers in software engineering leave the implementation details 
to the engineer, providing actionable feedback when required. 
Analogously, participants tended to provide less expert 
information to the agent up-front, and more as feedback;
they provided expert information in 49\% of prompts 
that asked the agent to generate a new code change, compared to 
in 66\% of prompts that ask the agent to refine a code change (i.e., iterate upon a previously generated change).
This relationship was even stronger for the 12 participants in 
managerial roles -- they provided expert insights in 45\% of their
prompts seeking new changes versus 75\% when seeking to refine changes.
As shown in Figure~\ref{fig:perception},
the analogy is reflected in the post-study questionnaire, 
in which participants report viewing the agent more as an 
assistant (at a lower level in the organizational hierarchy) 
than as a colleague (at the same level) 
or advisor (at a higher level).

While all participants usually tended to 
provide expert information in feedback prompts rather than up-front, 
we found that experience factors moderated how often participants provided such information.
For example, participants who had more familiarity with the repository tended 
to provide expert information more often -- those with 
over 10 commits to the repository in the four months leading up to the 
study provided such insights 70\% of time when seeking any code changes 
as compared to 48\% for other participants. 
Similarly, participants who had previous experience with Cursor also 
provided more expert insights, doing so in 81\% of prompts seeking 
code changes. 
Further, participants applying the one-shot strategy provided expert insights less often (in 47\% of their prompts seeking code changes) as compared to those applying the incremental resolution strategy (who did so in 64\% of their prompts seeking code changes).

\subsubsection{Requested Tasks}

We observed that participants
not only requested the SWE agent to make code changes to resolve the task specified in the  GitHub issue but also requested assistance in sub-tasks corresponding to other activities of the software development process. 
While 50\% of participants' prompts requested code changes, 
27\% of their prompts sought explanations of details related to the existing codebase (code comprehension), 
16\% of their prompts sought to run tests and 
11\% of their prompts sought explanations related to changes made by 
the agent (code review).
Note that the above categories are not mutually exclusive.
This distribution of prompts reflects participants' perceptions of 
the agent as reported in the post-study questionnaire (Figure \ref{fig:perception}). 
Participants viewed the agent the most as a content generator; 
prompts requesting code generation were most common. 
Since `reference guide' was a close second, 
participants also often used prompts to ask the agent about the codebase. 
On the other hand, participants did not view the agent as a reviewer, 
and were thus less likely to ask it to run tests or review its own code.


\subsubsection{Manual Actions} \label{manact}

Participants' perceptions of the agent shaped not only the types of 
prompts they provided the agent, 
but also the role the participants themselves took in the collaboration,
tending to work on debugging and testing more than on writing code.
Since participants reported not thinking of the agent 
as a code reviewer (Figure~\ref{fig:perception}), 
they often took the responsibility to debug and test code upon themselves,
doing so manually in 21 out of 33 issues by running tests or examining execution logs. 
In the case of bugs concerning UI changes, 
it was also easier for the participant to test changes manually by 
simply observing the state of the UI rather than uploading images to the chat for the agent to interpret.

On the other hand, since participants tended to view
the agent more as a content generator, 
they were more likely to let the agent handle all the code 
changes to be made, manually writing code in only 14 issues. 
Further, in 10 of these issues, participants' code interventions 
were limited to editing code suggested by the agent 
rather than adding new functionality.



Some participants expressed desire for AI tools to provide more 
support in non-code generation actions such as localizing, 
debugging, and testing, since these aspects of the software 
development process were often critical to resolving issues. 
P12 expressed that that the agent was not particularly helpful 
in localizing their issue, mentioning that 
\textit{“I think the main part [of solving the issue] was 
finding where to put the code, so it might have been marginally 
faster to do it myself”}. 
Similarly, participants mentioned the challenges they faced when 
collaborating with the agent for debugging; for instance, P8 responded 
in the post-study questionnaire that the agent was 
\textit{``not too great or useful at the debugging and converting-a-repro-into-a-test loop"}, 
and P11 responded that \textit{``It's puzzling to me why [the agent is] 
so good at generating code but bad at dealing with basic 
environmental commands"}. 
Participants' manual actions thus reflected not only their perceptions of the agent, but also a forced response to the constrained 
capabilities of the agent as a collaborator.

\subsubsection{Reviewing Agent Outputs}

The agent provided three types of outputs to communicate with participants:

\begin{itemize}
    \item\textit{Execution}: a trace of the agent's work that appears
    while the agent is working, including details like files read and commands executed
    (the bulk of Figure~\ref{fig:workflow}).
    \item\textit{Change Explanation}: at the end of a trace, typically the agent summarized
    the change it made to the participant's codebase (Figure~\ref{fig:workflow}c).
    \item\textit{Change Diff}: inside the participant's editor, the agent provides
    a number of inline diffs as proposed changes that the participant can accept or reject (Figure~\ref{fig:additional}).
\end{itemize}

We observed that participants tended to first track the agent's execution live 
as-it-happened before reviewing explanations and diffs, 
with a preference for reading diffs. 
While Cursor Agent allows for users to synchronously perform manual tasks in parallel to its
own execution,  
participants instead followed the agent's execution as it 
happened in real time for 84\% of their prompts.
Usually, this live analysis was sufficient for participants 
to understand the agent's \textit{process}; after following the agent's execution live, participants only reviewed the process that the agents followed (such as searching the codebase and reading files, in Figure~\ref{fig:workflow}b) for 7\% of the agent's responses.
In contrast, for the majority of cases (specifically, after 75\% of the agent's responses), 
participants reviewed the agent's \textit{outputs} (diffs or explanations) after following its execution in real time. 

\begin{figure}[t]
    \centering
    \includegraphics[width=\linewidth]{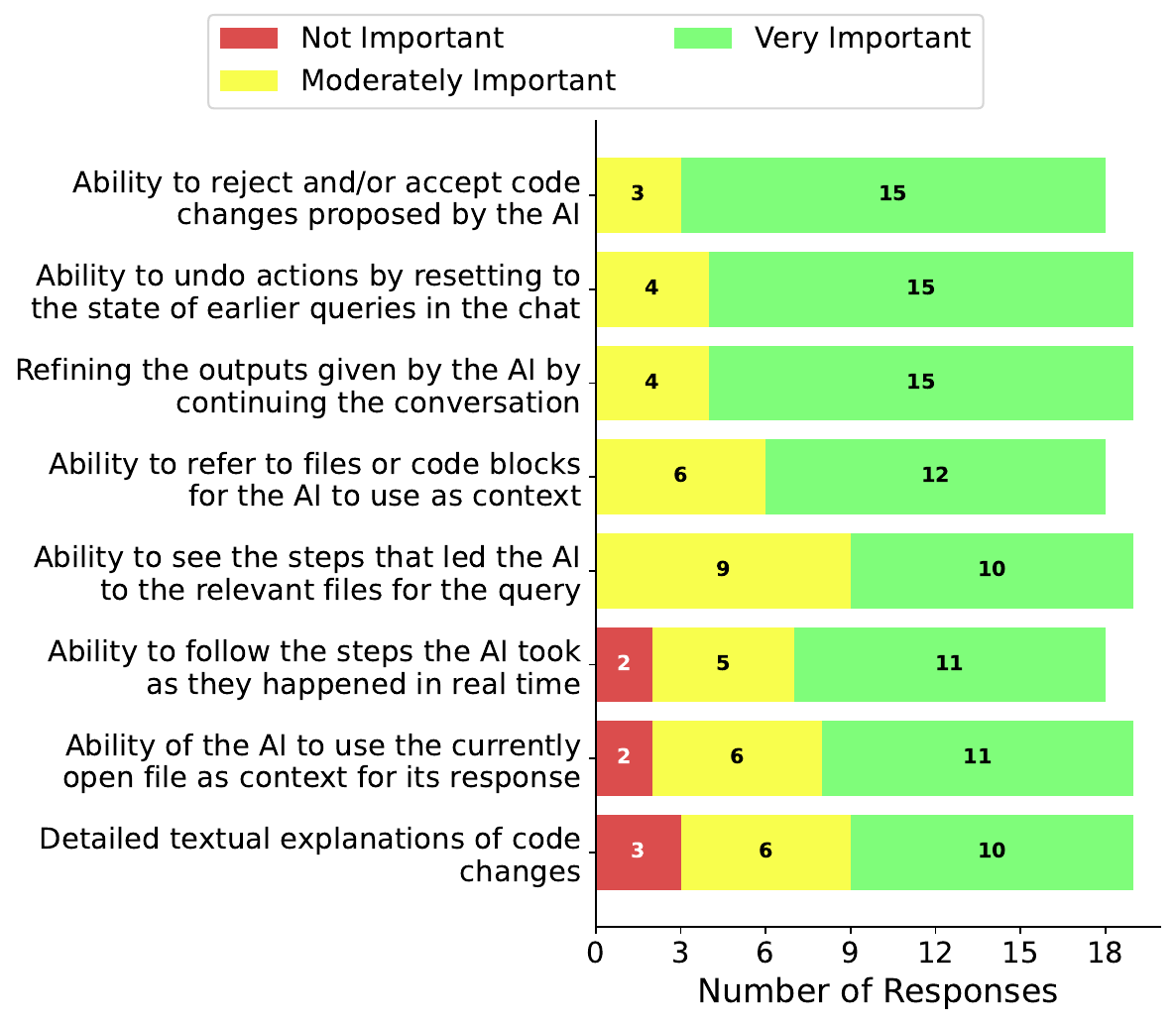}
    \vspace{-0.65cm}
    \caption{Post-study questionnaire responses on Agent Features}
    \label{fig:featuresurvey}
    \vspace{-0.4cm}
\end{figure}

When reviewing code outputs, participants preferred to 
verify diffs over change explanations.
While participants followed the agent's execution in 
real time for 95\% of the responses that included code changes, 
these code changes often required further verification -- participants 
reviewed diffs after 67\% of such agent responses, 
while they only reviewed explanations after 31\% of these responses. Only for 8\% of responses with code changes did participants review code explanations without reviewing code diffs.
This preference to review code directly is also reflected in the post-study questionnaire, 
in which participants ranked the ability to see detailed explanations of 
code changes as the least important feature among the different 
features of the agent (Figure~\ref{fig:featuresurvey}). 
Manual verification of the agent's outputs was usually sufficient for participants to 
have clarity on the agent's code changes -- they only asked for explanations 
of generated changes in 16\% of their follow-up prompts.


Participants who had previous experience with Cursor 
tended to focus even less on explanations than their peers, 
without any apparent impact on their success (Section~\ref{priorsuccess}). 
Like other participants, they followed the agent during 95\% of responses 
containing a code change, and further verified diffs in 
66\% of these responses. 
However, they only read explanations after 
18\% of agent responses with code changes, as 
compared to 35\% for other participants.

\subsubsection{Iteration Patterns}
Participants in our study often iterated on the agent's outputs 
to get to a satisfactory solution, 
preferring to refine faulty patches rather than starting over. Iteration was  required more often when working on 
feature requests -- participants used an average of 11.2 prompts when working on feature requests as compared to 6.3 when working on bugs.
Overall, participants used an average of 8.2 prompts per issue, 
out of which 
52\% of prompts that sought code changes sought refinements to 
previous changes by the agent, indicating that the agent's first outputs were often unsatisfactory and 
required further modifications. 
Despite this, participants only rejected code changes 10\% of 
the time after the agent generated code changes. 
Further, even though the agent allowed for participants to reset the chat
to earlier points in the 
conversation (Figure~\ref{fig:workflow}a) -- a feature we mentioned in the warm-up task -- participants only used 
this feature in 15\% of the issues they worked on. 
Additionally, participants only terminated the agent's execution 
before completion (Figure~\ref{fig:workflow}d) in 11\% of prompts. 
Thus, participants preferred to iterate on imperfect changes rather than starting afresh by stopping or resetting the agent. 

Despite this proclivity to iterate on imperfect outputs,
some participants indicated that they did not think that iterating 
was always a good strategy. P6 noted that the agent was \textit{``not understanding the implications of previously written bad code, and doubling down by adding more code instead of understanding the underlying issue."}
Similarly, P14 mentioned of the agent in their post-study questionnaire that 
\textit{``if you continue to ask it to fix issues that it hallucinated 
on you might end up wasting time following in on the wrong path. 
It would be better to just explain the task in more detail or 
revert back to a previous breakpoint."}

\subsection{RQ2: Barriers to Developer-Agent Collaborations}

\subsubsection{Lack of tacit knowledge} \label{tacit}
The agent was sometimes an ineffective collaborator 
because it lacked \textit{tacit} knowledge,
which is gained through personal experience and is undocumented~\cite{Rus_tacitk}.
Unlike with expert insights related to implementation details, it may be difficult for participants to know \textit{when} to discuss tacit knowledge, and even when they do, 
it may be difficult for them to articulate it in writing. Since the agent cannot understand non-verbal clues, this may hamper effective collaborations~\cite{haivshhwizardofozkuttal}.

For example, in the case of P7 (who had prior experience with Cursor), the participant was able to interpret a temporal cue but unable to communicate their finding to the agent -- when a terminal command running unit tests that was started by the agent seemed to be taking a long time to complete execution, the participant realized based on their experience with the repository that the tests were failing. However, rather than stopping the agent's execution and explaining this in a prompt, they chose to work on the issue manually while the agent came to this realization itself.
In other cases, the agent suffered from a lack of social context. 
For example, the agent once offered an explanation of the 
codebase that seemed to contradict comments on the 
issue by a maintainer of the repository,
making participant P8 skeptical of the agent's correctness. 
Prior knowledge of this context might have led the agent to 
question its assumptions before generating 
such an explanation.


\subsubsection{Unsolicited Actions}

We observed that the agent was sometimes excessively proactive, making changes beyond the scope of the participant's prompt.
We found that
the agent made unsolicited code changes after 38\% of participants' prompts that did not seek any code changes. 
In some cases, this was productive. 
For example, P5 mentioned that the agent \textit{``anticipated what I wanted to do".} 
In other cases, this led to miscommunication. 
For example, when the agent made changes 
beyond what the prompt specified, 
P16 was forced to reject the change and 
write a follow-up prompt to specifically instruct the agent 
to only make the requested change. 
P7 mentioned that the agent 
\textit{``always tries to generate new code, 
even though the right thing to do here might be [something else]"}. 

We further observed that the agent ran terminal commands in 
10\% of prompts that did not request to run any. 
As further evidence that these commands were unwanted,
participants were nearly three times more likely to stop the agent prematurely 
when its response included terminal commands (61\% vs. 21\%).
Unlike in code changes, where the agent does not apply proposed changes until 
explicitly accepted and can undo its changes, terminal executions can permanently 
change the state of the environment and the codebase. 
For example, while P12 was reviewing a test log file, 
the agent ran some new tests on the terminal, 
even though this was not instructed in the prompt. 
This led to the test log being deleted, 
derailing the participant's workflow. 
Similarly, P11 had to abandon an issue due to the 
agent corrupting the code environment 
by running a long series of terminal commands. 




\subsubsection{Sychophancy}
Sometimes, the tendency of the agent to agree with whatever the participant 
mentioned in their latest prompt led to a mistrust of the agent. 
In particular, after the agent produced a diff and an explanation for that diff,
participants were hesitant to trust the agent when in subsequent 
responses it offered a seemingly contradictory explanation for the same fix. 
As P15 mentioned, \textit{``If I say that 'your change is wrong', 
it will revert that change. That makes me lose confidence in it"}. 
Similarly, P16 expressed in the post-study questionnaire that the 
agent \textit{``can be confused easily and mislead the developer so [the agent] can't be trusted blindly"}. 
Bansal and colleagues~\cite{bansal2024challengeshumanagentcommunication} have also noted that consistency in the agent's outputs in critical to establishing trust in the agent.
P18 applied an alternate approach that avoided such issues, explaining that they 
\textit{``tend to treat [the agent] like a human, where I don’t want to 
give it the answer, I want it to think about it, and come to a 
conclusion itself. It might have insights I don’t know myself"}. 
They did this by asking the agent suggestive questions 
(e.g., \textit{are there any negative consequences of...}) 
rather than instructing the agent. 
This allowed the agent to 
retrospect while making refinements rather than blindly changing its 
assumptions based on the participant's latest instructions. 
This approach worked well for the participant -  they were able to solve the issue as well as find additional related problems 
with the codebase (unnecessary type checking) that they did not know about previously. 



\subsubsection{Overconfidence}
Kuttal and colleagues~\cite{haivshhwizardofozkuttal} report that AI agents for programming should be designed to respond with uncertainty when appropriate. 
Consistent with this finding,  we found in our study that participants sometimes mistrusted the agent 
due to its certainty that the changes it made in its latest response correctly solved the issue.
As P9 noted in the post-study questionnaire, 
\textit{``It never stopped thinking it knew how to solve the problem even though it was wrong"}. 
Even when the agent generated correct changes and explanations, 
its confidence in the alignment of these outputs to the issue sometimes 
led the participant to be hesitant of the agent. 
P18 noted about a generated fix 
that \textit{“it works, but it's not really what I want”}. 
Similarly, upon reviewing a description of codebase files that agent 
suggested the user to analyze, P10 mentioned, 
\textit{``These things are true, but not necessarily things I want to modify [to solve the issue]”}. 
Khati and colleagues~\cite{khati2025mappingtrustterrainllms} similarly report that AI trust metrics for developers include not only correctness, but also alignment with their intentions.

Participants were particularly hesitant when the agent made many changes. 
Participants stopped the agent's execution prematurely in 39\% of the 
responses in which the agent performed more than 3 
actions (such as terminal commands or code changes), as compared to 
only 9\% when the agent performed 3 or fewer actions. 
Participants sometimes expressed their concerns over losing control 
of the agent -- while P10 mentioned that \textit{``I don’t
feel super comfortable when it makes edits and I can't 
track exactly what and where"}, 
P14 noted about an issue that they were not able to 
successfully complete that \textit{``the problem I had with 
the agent is that it took over, I had to force delete things and it created a mess"}.
\subsection{RQ3: Factors Associated with Success} \label{greatsuccess}

We considered participants to be successful in solving the issue if they were completely satisfied with the fix so as to submit it as a pull request (\textit{Complete Success}, 10 issues) or if they were satisfied with the generated patch but needed to run some configuration steps or tests outside the scope of the study session before submitting the change as a pull request  (\textit{Patch Success}, 6 issues).
Participants made progress but were unable to fully complete 10 issues (\textit{Incomplete}), did not make any progress in 2 issues (\textit{No Progress}), and abandoned 1 issue due to an inability to make any progress towards the fix (\textit{Abandon Fix}).
Although the total number of issues that participants worked on (N=33)
is too low to conclusively reason about participants' success in resolving issues, we observed two main factors associated with success. We summarize the effect of these factors in Table \ref{tab:results}, and describe them in detail below.

\subsubsection{Collaboration Strategies} \label{successcollab}
Participants were more likely to be successful in collaborations 
with the agent when they took a more active role in this collaboration.

For example, participants were more likely to successfully 
resolve issues when they communicated more with the 
agent -- the average number of prompts used for successful 
issues was 10.3 as compared to 7.1 for unsuccessful issues. 
Iterating on the agent's outputs was also a factor in 
success -- 
participants were only successful in 
30\% of the issues in which they did not ask for code refinements.

Notably, participants who applied the one-shot strategy (Section~\ref{sec:collab-sat}) were 
less successful in solving issues -- they only succeeded in 38\% of the 
issues for which they provided the agent with the entire issue description. 
On the other hand, participants who applied the incremental resolution 
strategy were successful in 83\% of the issues they worked on. 
Similarly, participants who provided the agent with expert insights (Section~\ref{sec:expert})
tended to perform better than those who simply provided the agent with 
environmental context. Participants succeeded in 14 out of 22 (64\%) issues in which
they supplied such expert insights to the agent, as compared to only 2 out of the 7 (29\%) issues in which they did not provide 
any expert knowledge. 

Beyond providing insights to the agent, 
we also found that participants were more successful 
when they manually wrote code rather than just relying 
on the agent to do so. 
Participants were successful in 73\% of the issues in which 
they wrote code manually, as compared to in 36\% of the issues 
for which they did not write any code manually. 
This result did not extend to manual debugging and testing -- participants 
who manually worked on debugging or testing code for an issue were successful 
in 53\% of such issues, as compared to 60\% for issues in which they did not 
perform manual debugging or testing. 


\begin{table}[t]
\caption{Participant Success vs Study Characteristics}
\centering
\begin{tabular}{@{}p{2.5cm}p{4.5cm}p{1cm}@{}}
\toprule
\textbf{Factor} & \textbf{Attribute} & \textbf{Success Rate} \\
\midrule
\multicolumn{3}{@{}l@{}}{\textbf{Collaboration Strategies}} \\
\midrule
\multirow{2}{*}{Delegation Strategy}  & One-Shot & 38\% \\ 
& \textbf{Incremental Resolution} & \textbf{83\%}\\

\arrayrulecolor{gray!50}\cmidrule{1-3}\arrayrulecolor{black}
\multirow{2}{*}{Manual Actions} & \textbf{Manual Code Writing/Editing} & \textbf{73\%}\\  
 & No Manual Code Edits & 36\% \\ 

\arrayrulecolor{gray!50}\cmidrule{1-3}\arrayrulecolor{black}
\multirow{2}{*}{Insights Provided} & \textbf{Expert Insights} & \textbf{64\%}\\  
 & No Expert Insights & 29\% \\ 

\arrayrulecolor{gray!50}\cmidrule{1-3}\arrayrulecolor{black}
\multirow{2}{*}{Iteration Pattern} & \textbf{Requested Code Refinements} & \textbf{68\%}\\  
 & No Code Refinements Requested & 30\% \\ 

 \arrayrulecolor{gray!50}\cmidrule{1-3}\arrayrulecolor{black}
\multirow{2}{*}{Conversation Length} & Less than 10 prompts  & 50\%\\  
 & \textbf{10 prompts or more} & \textbf{64\%} \\ 
\midrule
 \multicolumn{3}{@{}l@{}}{\textbf{External Factors}} \\
\midrule

\multirow{2}{*}{Task Type} & Bugs & 38\%\\  
 & \textbf{Other Issues (Features, UI Fixes, etc.)} & \textbf{69\%}\\ 

\arrayrulecolor{gray!50}\cmidrule{1-3}\arrayrulecolor{black}
 \multirow{2}{*}{Experience with Cursor} & \textbf{Had Prior Experience} & \textbf{75\%}\\  
 & No Prior Experience & 52\%\\ 


\bottomrule 
\end{tabular}  
\label{tab:results}  
\vspace{-0.5cm}
\end{table}

\subsubsection{External Factors}\label{priorsuccess}

Participants' success was influenced by external factors relating to the tool and the task, such as prior experience and nature of the issue. The 3 participants with previous experience with Cursor had more success (3 out of 4 issues resolved), compared to others' 52\% success-rate.

Participants were most successful solving bugs that only 
concerned user interface changes (5 out of 5 issues resolved), 
followed by refactoring/variable renaming issues (2 out of 3 issues resolved), 
then feature requests (4 out of 8 issues resolved), 
and finally bugs (5 out of 13 issues resolved). 
These results suggest that participants may have struggled to 
collaborate with the agent to work on steps specific to 
bug resolution such as localization and debugging.



While benchmarks suggest that autonomous agents 
perform differently when working on code in 
different programming languages~\cite{zan2025multiswebenchmultilingualbenchmarkissue}, 
our human-in-the-loop study suggests that this was not an important factor 
in participants' success.
Participants had similar success rates -- succeeding 
about half the time -- across different languages (Python: 4 of 9 resolved successfully; 
TypeScript: 5 of 8; C++ 3 of 6; Cmake: 1 of 2; and Java 0 of 1), with the 
exception of C\# (3 of 3).


\section{Discussion}


\subsection{Collaborating Actively across the Development Process}
A recent blog post by Anthropic~\cite{anthropic2025} discusses the prevalence 
of the `feedback loop' interaction pattern, in which users collaborate 
with agents to work on code tasks simply by providing the agent with 
feedback from the environment. 
In our study, we observed a similar interaction pattern, as many of participants' 
prompts requesting code changes consisted only of environmental context or 
details of the original issue description. 
However, our results suggest that such interactions are less effective 
than those in which developers provide their own insights to the agent (Section \ref{successcollab}). 
Thus, when collaborating with agents, developers should not restrict 
their role to `human routers' \cite{knowledgeatwharton2016} that simply provide environmental feedback, but rather actively collaborate with them to achieve better results. 
Participants in our study often took a managerial role in these collaborations, which suggests that agents may be more likely to perform the role of junior engineers in developer-agent software teams. This in turn suggests that the junior developer workforce may be more impacted by the introduction of AI into software development teams.
Future work can explore how SWE agents can adopt the communication strategies and behaviors exhibited by high-performing subordinate engineers, and whether adopting these strategies affects developers' perceptions of such agents.

Our results further suggest that developers should actively collaborate with the agent in all stages of the software engineering process, not only in those stages for which the agent is not particularly helpful, such as debugging and testing.
Even though participants primarily used the agent to generate code changes, 
they were more successful when they collaborated with the agent by also manually writing code. On the other hand, even though participants expressed misgivings about the 
agent's ability to help with debugging (Section \ref{manact}), manually debugging and 
testing code did not appear to improve their likelihood of success. 


Thus, developer-agent collaborations were most successful when both the developer and the agent actively contributed to issue resolution.
In line with prior findings~\cite{steeringmultiagent}, as the agent was not as supportive in debugging and localizing, it was often unable to actively contribute to this part of the issue resolution process, and thus
participants struggled to successfully resolve bugs.
While software engineering benchmarks have clear, 
well set up tasks and environments for agents to work on, 
real development can involve messy environmental setup and 
debugging steps before the developer is ready to make code changes. 
However, agents often fail at these tasks, as reflected in their 
worse performance on more realistic benchmarks~\cite{miserendino2025swelancerfrontierllmsearn}.
For SWE agents to be useful for developers in their day-to-day tasks, 
they must empower users across all the steps of the software engineering process, 
especially the ones that are messy and ambiguous.

\subsection{Calibrating Output Scope}
AI agents are designed to solve complex issues autonomously, and have the tools to do 
so~\cite{liu2024largelanguagemodelbasedagents}. 
However, we found that this design can backfire -- in our study, participants who 
applied the one-shot delegation strategy were often unsuccessful (Section \ref{successcollab}).
When the agent tries and fails to successfully implement complex changes, 
it is difficult for the developer to debug where the problems in the buggy patch arise from. 
Further, real-world complex issues often require expert information that only an experienced developer might know. 
Yet, knowing what information could be relevant may be difficult for developers to foresee, and providing all such necessary information in a single prompt might be 
tedious for developers, particularly in the case of complex issues. 
Thus, interactive agents should leverage the presence of the developer as a collaborator by internally calibrating the scope of their outputs when working on an issue. 
This could enable the developer to provide expert insights when needed, 
and effectively investigate any problems with the agent's output. Bajpai and colleagues~\cite{robin} similarly found that in-IDE AI copilots 
are more effective when debugging if they focus on gathering information and 
communicating with the user before suggesting a fix. 
Such a pipeline, as also supported by \cite{guidelineshaichi2019amershi}, has the advantages of the incremental resolution strategy 
without requiring additional effort from the developer to manually analyze the issue or intervene in the fix unless necessary.


\subsection{Collaborating through Discussion, Not Sycophancy}

Friction arising from the agent's lack of tacit knowledge is
difficult to anticipate for both the agent and
the developer (Section~\ref{tacit}). 
This friction may occur regardless of how much
context the agent gathers, 
meaning that agents may always produce incorrect patches in some cases. 

In contrast to pair programming, where developers spend 
considerable time discussing code logic and implementation details~\cite{Prechelt2009, haivshhwizardofozkuttal}, 
the agent rarely engaged in problem-solving discussions. 
Instead, despite initial confidence in its changes, 
the agent immediately agreed with whatever 
the developer said in the latest prompt. 
While participants using incremental strategies were more successful, the agent contributed little to the solution quality since it simply followed the user's instructions.

However, when participant P18 invited the agent to 
challenge their opinions, they uncovered novel insights about the existing codebase,  a synergy highlighted in recent surveys of LLM-backed SE agents~\cite{liu2024largelanguagemodelbasedagents}.
This suggests that agents designed to challenge rather than merely obey the developer -- by engaging in substantive discussion -- might not only increase trust and improve solution quality, but further promote developer growth through critical thinking~\cite{challenge}.

\section{Conclusion}
In this paper, we've seen that developers can collaborate with 
a SWE agent to solve real-world software engineering
tasks, but also that such a collaboration is fraught
and that success is not guaranteed.
While the capabilities of large language models and of the 
agents that build upon them expands at a dizzying rate, 
the capabilities of the humans who work with them are relatively
static.
The strategies, challenges, and interaction patterns we 
document here are therefore likely to remain relevant even as 
SWE agents rapidly evolve. 
By understanding and addressing these enduring human factors, 
we hope this work contributes to a future where developer-agent collaboration is both effective and empowering.

\bibliographystyle{ieeetr}
\bibliography{references} 

\appendix \label{appendixx}

\subsection{Qualitative Codebooks}
Tables \ref{tab:activitytaxo} and \ref{tab:trajectorytaxo} show the details of our codebooks for coding participant actions and chat trajectories respectively.

\begin{table*}[!htbp]
\caption{Participant Activity Taxonomy}
\centering
\begin{tabular}{@{}p{4.5cm}p{12.5cm}@{}}
\toprule
\textbf{Activity} & \textbf{Description} \\
\midrule
\multicolumn{2}{@{}c@{}}{\textbf{IDE Actions}} \\
\arrayrulecolor{gray!50}\cmidrule{1-2}\arrayrulecolor{black}
\rowcolor{gray!5}
Checkpoint & Revert to an earlier checkpoint in the conversation \\
Accepted Code Change & Accept code changes proposed by the agent \\
\rowcolor{gray!5}
Rejected Code Change & Reject code changes proposed by the agent \\
Stopped Agent & Stop agent run prematurely \\
\addlinespace[0.3em]
\multicolumn{2}{@{}c@{}}{\textbf{Communication}} \\
\arrayrulecolor{gray!50}\cmidrule{1-2}\arrayrulecolor{black}
\rowcolor{gray!5}
Prompt Crafting & Writing a prompt in the chat window \\
Prompt Submitted & Submitting a prompt in the chat window \\
\rowcolor{gray!5}
Following Agent Live & Waiting for/ following the steps of agent while it runs \\
Thinking/Verifying Explanations & Actively thinking about and/or verifying textual information by the agent \\
\rowcolor{gray!5}
Thinking/Verifying Actions & Actively thinking about the actions taken by the agent to reach its final output (e.g. searching codebase for relevant files) \\
Thinking/Verifying Code & Actively thinking about and/or verifying a proposed or accepted code change by the agent \\
\addlinespace[0.3em]
\multicolumn{2}{@{}c@{}}{\textbf{Code Actions}} \\
\arrayrulecolor{gray!50}\cmidrule{1-2}\arrayrulecolor{black}
\rowcolor{gray!5}
Writing New Code & Manually writing code that implements new functionality \\
Debugging/Testing Code & Manually running or debugging code to check functionality (e.g. testing UI changes on local browser) \\
\rowcolor{gray!5}
Editing Suggested Code & Manually editing code changes suggested by the agent \\
Editing (Personally) Written Code & Editing code written by a programmer that is not a suggestion by the agent for the purpose of fixing existing functionality \\
\addlinespace[0.3em]
\multicolumn{2}{@{}c@{}}{\textbf{Non-Code Actions}} \\
\arrayrulecolor{gray!50}\cmidrule{1-2}\arrayrulecolor{black}
\rowcolor{gray!5}
Looking up issue & Referring to issue description/ details \\
Looking up Documentation & Checking an external source for the purpose of understanding code functionality or other information needed to help work on the issue (e.g. Stack Overflow) \\
\rowcolor{gray!5}
Waiting for Terminal & Waiting for terminal executions to complete \\
Deferring Thought For Later & Programmer accepts suggestion without verifying/understanding it, but plans to verify it after \\
\rowcolor{gray!5}
Thinking About New Code To Write & Thinking about what code or functionality to implement and write (e.g. reading/navigating through the existing codebase) \\
\bottomrule
\end{tabular}
\label{tab:activitytaxo}
\end{table*}

\begin{table*}[!htbp]
\caption{Chat Trajectory Taxonomy}
\centering
\begin{tabular}{@{}p{3.5cm}p{13.5cm}@{}}
\toprule
\textbf{Code} & \textbf{Description} \\
\midrule
\multicolumn{2}{@{}c@{}}{\textbf{Information Given to Agent in Prompt}} \\
\arrayrulecolor{gray!50}\cmidrule{1-2}\arrayrulecolor{black}
\rowcolor{gray!5}
Issue Knowledge & Providing agent with instructions/knowledge about the issue based solely on original issue description, if any \\
Implementation Knowledge & Providing agent with instructions/knowledge about the issue related to implementation/configuration/scope etc., not solely based on original issue description or environmental context \\
\rowcolor{gray!5}
Styling Knowledge & Providing agent with instructions/knowledge about the issue related to code styling/practices not solely based on original issue description or environmental context \\
Affirmation & Affirming the follow up provided by the agent \\
\rowcolor{gray!5}
Meta-details & Informing the agent of plan-level heuristics or steps to follow in its response, not grounded in the codebase \\
\addlinespace[0.3em]
\multicolumn{2}{@{}c@{}}{\textbf{Response Sought from Agent in Prompt}} \\
\arrayrulecolor{gray!50}\cmidrule{1-2}\arrayrulecolor{black}
\rowcolor{gray!5}
Change Explanation & Seeking explanations of code changes made by the agent \\
Codebase Explanation & Seeking explanations pertaining to the existing codebase \\
\rowcolor{gray!5}
External Explanation & Seeking explanations pertaining to external knowledge not rooted solely in the codebase \\
Refine Change & Seeking corrected version of old code change \\
\rowcolor{gray!5}
New Change & Seeking new code changes \\
Testing & Seeking to test existing/previously generated code \\
\addlinespace[0.3em]
\multicolumn{2}{@{}c@{}}{\textbf{Agent Actions}} \\
\arrayrulecolor{gray!50}\cmidrule{1-2}\arrayrulecolor{black}
\rowcolor{gray!5}
New code change & Add code changes to a file \\
Refine code change & Refine previously made changes to a file \\
\rowcolor{gray!5}
Run terminal command & Run terminal command \\
Explain plan & Textual description of multi-step plan to be performed by the agent \\
\rowcolor{gray!5}
Explain codebase & Explains details of existing code in the codebase \\
Explain old changes & Provides textual explanations of code changes made in previous runs \\
\rowcolor{gray!5}
Explain new changes & Provides textual explanations of code changes made in current run \\
Search Web & Searches the internet \\
\rowcolor{gray!5}
Explain external & Explains external details (usually based on web search) \\
Search codebase & Searches through the codebase \\
\rowcolor{gray!5}
Analyse files & Reads/analyses files in the codebase \\
Follow Up & Prompts the participant for follow-up queries (usually at end of response) \\
\bottomrule
\end{tabular}
\label{tab:trajectorytaxo}
\end{table*}

\subsection{Additional Results}
In this section, we describe some other observations we made during the study regarding barriers to effective developer-agent collaborations that we did not 
have space to include in the main paper.

\subsubsection{Ineffective Follow-Up Suggestions}
The agent offered textual suggestions for follow-up prompts 
in 62\% of its responses.
For participants, these suggestions were usually 
not effective in advancing the conversation for participants, 
as participants  replied to only 7\% of such follow ups 
in their next prompt. 
One reason for this may be that these suggestions were often 
non-specific (e.g.:  \textit{Would you like me to make any 
additional changes or would you like me to explain any part of the solution in detail?}).
When participants did reply, the suggestions were more 
specific (e.g.: \textit{Would you like me to show you how to 
remove the manual tracking code and let RichEdit handle double-clicks naively?}).
Further, while participants often needed refinements 
to code changes, the agent's follow ups instead would often suggest 
moving forward to other steps in the issue resolution process. 
Thus, participants may also not have 
responded to follow ups because they
did not always reflect the true state of the solution.

\subsubsection{Synchronicity}
Parallel actions by the agent and the user sometimes led to challenges 
stemming from the participant's unawareness of the actions being performed 
by the agent. 
While in theory simultaneous collaborations on the same codebase 
allowed for participants to never be blocked from directly 
working on the code, this did not always work out in practice. 
When P9 tried to work on code alongside the agent, it led to the 
agent duplicating a part of the fix, 
forcing the participant to reject the changes. 
To avoid such situations, some participants preemptively 
prevented themselves from working on the codebase.
P10 expressed that \textit{“I want to make sure that the code changes 
are clear and I can see the diffs clearly so I'll wait for it to stop thinking”}. 
While the agent was iterating on a fix and seemed to be going 
down the wrong path, P11 noted similar concerns: 
\textit{``It is doing something which is no longer helpful... 
I am scared it will break something, and so I will wait for it to finish what its doing"}.

\subsubsection{Verbosity}
Participants often sought and read explanations by the agent in 
their prompts, doing so after 39\% of the agent's responses. 
However, several participants (P6, P7, P13, P15, P18) expressed 
frustration with the verbosity of such explanations. 
While P15 noted that \textit{``Most of the time when I’m coding, 
I don’t need detailed responses, only to see the changes made”}, 
P16 mentioned of a response that 
\textit{``It's giving me a lot of gibberish... I don't need all this analysis"}. 
Sometimes, verbosity even led to miscommunication. 
In the case of P15, due to the verbosity of the agent's response, 
they did not notice that the response contained a documentation 
link that they were looking for, 
and instead searched the web manually for the link. 
Thus, while participants engaged with the agent's explanations, 
these explanations did not always provide participants with the 
level of details they needed for effective communication.

\subsubsection{Control}
In the post-study questionnaire, participants reported that the ability 
to accept and reject proposed changes by the agent and the ability 
to reset to earlier points in the conversation were the first and 
second (tied) most important features of the tool. 
This is despite the fact that participants did not often use these features -- they 
rejected code changes only 10\% of the time after the agent generated a code change, 
while they reset to earlier points in the conversation in only 5 out of the 33 issues 
they worked on. 
Thus, even though participants did not exercise control over the agent often, 
it was important for them to have a sense of control over the changes the agent made. 
P10 reaffirmed this feeling, 
noting \textit{``that’s the main problem with AI, even if it's correct, we can't trust it blindly"}.

\subsection{Additional Discussion}

\subsubsection{Dynamic Response Context}
Participants in our study almost always followed the agent 
in real time during its execution, 
and often read the explanations it produced. 
However, they also expressed frustration over the verbosity 
of the agent's responses. 
To solve this problem, agents could dynamically create summaries 
of their prior actions and explanations and only display in 
full length the final or latest code output or 
corresponding explanation. 
This would let developers understand relevant actions and 
explanations of the agent fully while it executes 
in real time, while not having to navigate though a 
long response when looking through the agent's final outputs. 
On the other hand, if the user wishes to look through 
full-length explanations, the agent could provide these as well. 
Allowing the user to exercise their discretion in viewing the agent's 
output could provide for a less frustrating experience 
while not sacrificing communication quality.

\subsubsection{Maintaining Live User Context}
Often, participants in our study submitted prompts to the 
agent that did not request code changes. 
However, the agent would sometimes proactively make changes anyway, 
assumedly driven by its goal to complete the task. 
This was helpful in some cases, but could also cause the user 
to be wary of the agent, potentially blocking them from 
working on the issue in parallel due to uncertainty about the 
agent's actions. 
To solve this issue, the agent could time its actions to the context 
of the user~\cite{guidelineshaichi2019amershi}. 
The agent can achieve this by maintaining a live context of the 
user's actions and deferring to the user when a conflict arises. 
For example, if a user is editing a file, the agent should stop making 
edits to that file until the user moves on to another action. 
Similarly, if the user is running tests or looking at execution logs, 
the agent should not run disruptive commands.
This can increase the user's confidence in the agent and encourage 
them to work synchronously with the agent without worrying about 
negative repercussions.

\subsection{Pre-Study Questionnaire}


\begin{enumerate}[label=\arabic*.]  
    \item With which gender do you most identify? \begin{itemize}  
        \item Woman  
        \item Man  
        \item Non-binary / gender diverse  
        \item Prefer not to say  
        \item Other  
    \end{itemize}  
  
    \item What is your age?  
    \begin{itemize}  
        \item 18-25  
        \item 26-35  
        \item 36-45  
        \item 46-55  
        \item 56-65  
        \item 66+  
        \item Prefer not to say  
    \end{itemize}  
  
    \item With which race/ethnicity do you most identify? (Select all that apply)  
    \begin{itemize}  
        \item American Indian or Alaska Native  
        \item Asian   
            \item Central Asian  
            \item East Asian  
            \item South Asian  
            \item Southeast Asian   
        \item Black or African American  
        \item Native Hawaiian or Pacific Islander  
        \item White  
        \item Hispanic or Latino  
        \item Multiracial  
        \item Prefer not to say  
        \item Other  
    \end{itemize}

    \item Which best describes your programming experience?\\  
    \textit{Scale: Multiple Choice}  
    \begin{itemize}  
        \item 0 to 2 years professional programming experience  
        \item 3 to 5 years professional programming experience  
        \item 6 to 10 years professional programming experience  
        \item 11 to 15 years professional programming experience  
        \item More than 16 years professional programming experience  
    \end{itemize}  
  
    \item When coding, how frequently do you use the following tools to \textbf{chat with AI}?  
    
    \textit{Scale: Never, Occasionally, Regularly}
    \begin{itemize}  
        \item ChatGPT  
        \item GitHub Copilot Chat  
        \item Codeium Windsurf  
        \item Cursor  
        \item Tabnine  
        \item Claude  
    \end{itemize}  
    (Scale for each: Never / Occasionally / Regularly)  
  
    \item If you have previously used any tools other than the ones mentioned above to chat with AI, please specify them below.   
    
    \textit{Open-ended response}

    \item When coding, how frequently do you use the following tools' AI-generated \textbf{code completions}?  

    \textit{Scale: Never, Occasionally, Regularly}
    \begin{itemize}   
        \item GitHub Copilot  
        \item Windsurf Cascade
        \item Cursor AI
        \item Tabnine  
        \item Blackbox AI  
    \end{itemize}  
  
    \item If you have previously used any AI tools other than the ones mentioned above for AI-generated code completions, please specify them below. 
    
    \textit{Open-ended response}

\item How frequently do you use AI developer tools?  
  
\textit{Scale: Never, Occasionally, Regularly}  
  
\begin{itemize}  
    \item to write new code  
    \item to write documentation for existing code  
    \item to modify existing code  
    \item to explain unfamiliar code  
    \item to debug code  
\end{itemize}  
  
\item Please rate your agreement with the following statements. AI tools for fixing bugs:  
  
\textit{Scale: Not at all, Moderately, Extremely, Don't Know}  
  
\begin{itemize}  
    \item are easy to use  
    \item can understand your requests  
    \item can provide high quality responses  
    \item can help you to write better code  
    \item can help you to write code more quickly  
    \item are enjoyable to use  
\end{itemize}  
\end{enumerate}  

\subsection{Post-Study Questionnaire}
\begin{enumerate}[label=\arabic*.]  

\item Please rate your agreement with the following statements. AI tools for fixing bugs:  
  
\textit{Scale: Not at all, Moderately, Extremely}  
  
\begin{itemize}  
    \item are easy to use  
    \item can understand your requests  
    \item can provide high quality responses  
    \item can help you to write better code  
    \item can help you to write code more quickly  
    \item are enjoyable to use  
\end{itemize}  
  
\item To what extent did you view the AI tool as:  
  
\textit{Scale: Not at all, A little, Somewhat, A great deal}  
  
\begin{itemize}  
    \item A reference guide  
    \item A content generator  
    \item A problem solver  
    \item A collaborator  
    \item A colleague  
    \item A coach  
    \item An advisor  
    \item An assistant  
    \item A reviewer  
\end{itemize}  
  
\item How important were these aspects of working with the AI tool?  
  
\textit{Scale: Not Important, Moderately Important, Very Important}  
  
\begin{itemize}  
    \item Detailed textual explanations of code changes  
    \item Ability to see the steps that led the AI to the relevant files for the query  
    \item Ability to follow the steps the AI took as they happened in real time  
    \item Ability to reject and/or accept code changes proposed by the AI  
    \item Refining the outputs given by the AI by continuing the conversation  
    \item Ability to undo actions by resetting to the state of earlier queries in the chat  
    \item Ability of the AI to use the currently open file as context for its response  
    \item Ability to refer to files or code blocks for the AI to use as context  
\end{itemize}  

    \item  Did anything stand out to you about the experience of using the AI tool? For example, was 
anything good, bad, surprising, or notable?
    
    \textit{Open-ended response}
    
\end{enumerate}

\subsection{Study Protocol}
Following is the study protocol containing details of the tutorial task and guidelines for the tasks participants performed that the study administrator followed across all sessions:
\vspace{0.5cm}

\textbf{Introduction}

Hi, good morning/afternoon/evening. Thanks for taking time out to participate in this user study. Before we begin, could you fill a short pre-study survey? I’ll send over the link to the form, and you can fill it on your system. Your participant number is \_\_\_\_

\textit{Participant fills the pre-study questionnaire}

Today, you will be using the AI in Cursor IDE by remotely controlling my system to work on issues in the \_\_\_\_ repository. You can choose to work on any issues that you feel you could solve by making direct edits to the codebase. The goal is for you to create a solution patch for the issue that you would feel confident attaching your name to and submitting as a pull request for review. If you get done solving the issue, we’ll choose another issue for you to work on. The session will be around an hour long, and you will fill out a short post-session survey at the end. As mentioned in the survey, I’ll be recording the meeting. Do you have any questions?

Great, let me share my screen.

\textbf{Warm-up Task}

Let’s go through a quick demo task so you are familiar with how the IDE and its AI features work. Cursor is a fork of VSCode, so it uses most of the shortcuts and functionalities present in VSCode. Today, we’ll be using the AI tool Cursor Agent. You can chat with Cursor Agent and ask it to implement code changes for you. Cursor Agent can read and analyse all any files in the IDE, make code edits, and also run terminal commands. Let’s try this out. I have cloned a GitHub repo (\href{https://github.com/Marfullsen/cli-ball-sort}{link}) for a CLI-based python game that involves sorting balls into different containers. Let’s try out this game on the terminal by running \texttt{python3 cli-ball-sort.py}. Great, so as you can see, there are 5 different containers and 3 balls to sort. So now, let’s say we want to use Cursor Agent to modify the game. On the chat window on the right, I’ve entered a prompt (\textit{Prompt entered: change the number of containers to 6, the maximum number of balls in each container to 5, and test out the changes}) that asks to change the number of containers to 6 and the number of balls to 5. Cursor uses the open file as context for its response, though it can also search through the codebase for other files as we’ll see. You can also manually add context to give to Cursor. For example, I can add some lines of this file (\texttt{customizer.py}) to the context. I can also remove context (\textit{remove those lines}). Let’s try giving it our prompt.

\textit{Cursor Agent runs and makes edits}

As you can see, the AI searched through the different files and made edits. You can now approve or reject these edits. Cursor has also run the game through the terminal, and it seems to be working.  Let’s check how it did by running it ourselves.

\textit{Run the game on terminal}

Great, since these changes look right, let’s approve them. Aside from rejecting these changes, you can also stop the agent at any point in its execution and restore to an earlier point in the conversation (\textit{demonstrate where this button is}). Do you have any questions about Cursor?

Great, let’s get started with the study. After you’re doing solving an issue, I will save the git diff of the changes you made and revert the repository back to the original state. I’ll email you these changes to you after the session. 

\textbf{Main Task}

The repository has been set up in Cursor along with the all the required installations on a new conda environment. As in the demo, you can use the terminal.

You are also free to access the internet. I have a browser window open with the open issues in this repository (\textit{open the browser window}), you can use this window to access any resources you need. 

 Please try to think aloud as you go though your process. Try to use the AI for all your tasks in the study, but feel free to make small manual changes if necessary.
 
Just to reiterate, your objective in this session is to solve the issue efficiently to the best of your ability, not to evaluate or observe how well the AI can solve the issue. 

Let me give you remote control of my screen. You can now look for an issue to start solving with Cursor Agent.

\textbf{Reminders}

Before moving on to a new issue, confirm if the participant is 100\% happy with the change, and they can attach their name on the patch when submitting a PR. If not, ask them to keep working on the issue till they get to this point. Only move on in this case if it is not possible for them to reach this level of satisfaction with the issue in the scope of the study.

Choice of issue:
\\1st issue: Entirely participant’s choice
\\Subsequent issues: Work with participants to choose issues varying in difficulty and domain (e.g. feature vs bug) depending on how successful the first issue was.

Prompts to remind the participant to think aloud:
\begin{itemize}
    \item Can you explain what you’re trying to do right now?
    \item Could you talk about what your thought process right now?
    \item What are you thinking about/ trying to do?
\item Tell me a little what you’re doing
\item Could you explain how what you’re doing is relevant to the task?
\end{itemize}

\textbf{Post-Task}

Thanks for participating! The session was very insightful. I’ll send over the link to the post-study survey so you can fill it out on your system.

\textit{Participant fills the post-study questionnaire}

Great! Before we end, do you have any questions? 

Okay, then once again, thanks for participating! Have a nice day!

\end{document}